\documentclass{article}
\usepackage{amsfonts,amssymb, amsmath}
\usepackage[english]{babel}

\def\nn{\nonumber }
\def\bq{ \begin{equation} }
\def\eq{ \end{equation} }
\def\ben{ \begin{eqnarray} }
\def\en{ \end{eqnarray} }

\def\ii{{\rm i}}

\newtheorem{re}{Remark}

\begin{document}


\title{Integrable systems on the sphere associated with  genus three algebraic  curves}
\author{ A. V. Tsiganov,  V. A.
Khudobakhshov \\
\it\small
St.Petersburg State University, St.Petersburg, Russia\\
\it\small e--mail:  andrey.tsiganov@gmail.com, vitaly.khudobakhshov@gmail.com}
\date{}
 \maketitle
\date{}
\maketitle

\begin{abstract}
New variables of separation for few integrable systems on  the two-dimensional sphere with higher order integrals of motion are considered in detail. We  explicitly describe  canonical transformations of  initial physical variables  to  the variables of separation and vice versa,  calculate the corresponding quadratures and  discuss some possible integrable deformations of initial systems .
\end{abstract}

\vskip0.1truecm
\section{Introduction}
\setcounter{equation}{0}
A fundamental requirement for new developments in mechanics is to unravel the geometry that
underlies different dynamical systems, especially mechanical systems.
There are several reasons why this geometrical understanding is fundamental. First, it is a key tool for reduction by symmetries and for the geometric characterization of the integrability and stability theories. Second, the effective use
of numerical techniques is often based on the comprehension of the fundamental structures appearing
in the dynamics of mechanical and control systems. In fact,  geometric analysis of such systems reveals what they have in common and indicates the most suitable strategy to obtain and to analyze their solutions.

Already in 19th century Euler and Lagrange established a mathematically satisfactory
foundation of Newtonian mechanics.  In \cite{jac42} Jacobi united  their ideas with the Hamilton optic theory
and with the Abel geometric methods at a new Hamilton-Jacobi formalism.
The Hamilton-Jacobi formalism was a crucial step towards Liouville's classical
definition of the notion of integrability \cite{li55}  based on the notion of first integrals of motion.

The  Liouville definition of integrable Hamiltonian systems naturally covered many classical examples.
Among them are the Kepler motion solved by Newton, harmonic oscillators
solvable by trigonometric functions, the Euler and Lagrange  spinning tops
and the Jacobi example of geodesic motion on an ellipsoid solvable by elliptic functions \cite{bm05},
the Neumann system on the sphere \cite{neu59} and Kowalevski top \cite{kow89} solved terms of
hyperelliptic functions etc. In novel times much attention owing to the another discovery of the vast
class of integrable soliton nonlinear partial differential equation, that admits
this type of integrability when dynamics is restricted to finite dimensional Liouville tori and the
system appeared to be completely integrable in the Liouville-Arnold sense.
They all are more or less connected with the hyperelliptic curves and with the hyperelliptic functions \cite{bbe94,jac42,st95}.
 Below we show that foregoing development of the theory detected a number of cases when associated algebraic curve is non hyperelliptic and and its genus exceeds the number degrees of freedom \cite{ts11s,ts09v,ts11v}.

Bi-Hamiltonian structures can be seen as a dual formulation of integrability and separability, in the sense that they substitute a hierarchy of compatible Poisson structures to the hierarchy of functions in involution, which may be treated either as integrals of motion or as variables of separation for some dynamical system \cite{ts07}. The Eisenhart-Benenti theory was embedded into the bi-Hamiltonian set-up using the lifting of the conformal Killing tensor that lies at the heart of Benenti's construction, which may be realized as a computer algorithm  \cite{tsg05}. The concept of natural Poisson bivectors allows us to generalize this construction and to study systems with quadratic and higher order integrals of motion in framework of a single theory \cite{ts07,ts10,ts11s}.

 The aim of this note is to discuss   separation of   variables  for  integrable natural systems on the two-dimensional unit sphere $\mathbb S^2$ from \cite{ts10c,ts10k,ts10,ts11s,ts09v}.  In the above mentioned  previous  papers we focused our attention on the bi-Hamiltonian calculations  of the variables of separation starting from the given integrals of motion. This note is devoted to construction of the initial physical variables in terms of variables of separation,
 to calculation of the corresponding quadratures and to discussion of the possible integrable "gyroscopic" deformations of these systems associated with genus three algebraic curves.

In order to describe integrable systems on the sphere we will use the angular momentum vector $J=(J_1,J_2,J_3)$ and the Poisson vector $x=(x_1,x_2,x_3)$ in a moving frame of coordinates attached to the principal axes of inertia. The Poisson brackets between these variables
\begin{equation}\label{e3}
\,\qquad \bigl\{J_i\,,J_j\,\bigr\}=\varepsilon_{ijk}J_k\,, \qquad
\bigl\{J_i\,,x_j\,\bigr\}=\varepsilon_{ijk}x_k \,, \qquad
\bigl\{x_i\,,x_j\,\bigr\}=0\,,
\end{equation}
may be associated to the Lie-Poisson brackets on the algebra $e^*(3)$.
Using the Hamilton function $H$ and the Lie-Poisson bracket $\{.,.\}$ (\ref{e3})
on  the Euclidean algebra $e^*(3)$ the customary Euler-Poisson  equations may be rewritten
in the Hamiltonian form
\bq\label{eq-mkov}
\dot{J}_i=\{J_i,H\}\,,\qquad \dot{x}_i=\{x_i,H\}\,.
\eq
Remind, that
the Lie-Poisson dynamics on $e^*(3)$ can be interpreted as resulting from
reduction by the symmetry Euclidean group $E(3)$ of the full dynamics on the
twelve-dimensional phase space $T^*E(3)$ \cite{bm05}. There are  two Casimir elements
\bq \label{caz-e3}
 C_1=|x|^2\equiv\sum_{k=1}^3 x_k^2, \qquad C_2= \langle x,J \rangle\equiv\sum_{k=1}^3 x_kJ_k,
\eq
where $\langle.,.\rangle$ means inner product.  Using canonical transformations $x\to \alpha x$ we will always put
$C_1=1$ without loss of generality.

 If the square integral of motion $C_2=\langle x,J \rangle$ is equal to zero, rigid body dynamics may be restricted on the unit sphere $\mathbb S^2$ and we can use standard spherical coordinate system on its cotangent bundle $T^*{\mathbb S}^2$
\bq\label{sph-coord}
\begin{array}{lll}
x_1 =\sin\phi\sin\theta,\qquad& x_2 = \cos\phi\sin\theta,\qquad & x_3 =\cos\theta\,,\\
\\
J_1 =\dfrac{\sin\phi\cos\theta}{\sin\theta}\,p_\phi-\cos\phi\,p_\theta\,,\qquad&
J_2 =\dfrac{\cos\phi\cos\theta}{\sin\theta}\,p_\phi+\sin\phi\,p_\theta\,,
\qquad& J_3 = -p_\phi\,.
\end{array}
\eq
We use these variables in order to determine canonical variables of separation  on $T^*\mathbb S^2$ .

As usual all the results are presented up to the linear canonical transformations, which consist of rotations
\bq\label{rot}
 x\to \alpha\, {U}\, x\,,\qquad J\to {U}\, J\,,
\eq
where $\alpha$ is an arbitrary parameter and $U$ is an orthogonal constant matrix,
and shifts
\bq
x\to x \,,\qquad J\to J+ {S}\, x\,,\label{shift}
\eq
where ${ S}$ is an arbitrary $3\!\times\!3$ skew-symmetric constant matrix.

Of course, any canonical transformation of the spherical variables (\ref{sph-coord}) yields
automorphism of $e^*(3)$ too.  For instance, trivial canonical transformation
\bq\label{theta-trans}
 p_\theta\to p_\theta+f(\theta)
\eq
gives rise to "generalized" shift depending on arbitrary function $f(x_3)$:
\bq\label{g-shift}
J_1 \to J_1-\dfrac{x_2f(x_3)}{\sqrt{x_1^2+x_2^2}}\,,\qquad
J_2 \to J_2+\dfrac{x_1f(x_3)}{\sqrt{x_1^2+x_2^2}}\,,
\eq
This and more complicated canonical transformations of $e^*(3)$ are discussed in \cite{bm05,kst03}.

\section{Kowalevski top and Chaplygin system}
\setcounter{equation}{0}
Following to \cite{ts10c,ts10k,ts11s}, we  determine  canonical coordinates $q_{1,2}$ on $T^*\mathbb S^2$ as  roots of the following  polynomial
\ben
B(\lambda)&=&(\lambda-q_1)(\lambda-q_2)
=\lambda^2-\dfrac{p_\theta^2\sin^2\theta+p_\phi^2\cos^2\theta}{\sin^\alpha\theta\cos^2\theta}\,\lambda-a^2-b^2
\nn\\
\label{kow-var}\\
&-&\dfrac{(a\cos\alpha\phi-b\sin\alpha\phi)(p_\theta^2\sin^2\theta+p_\phi^2\cos^2\theta)}
{\sin^\alpha\theta\cos^2\theta}
-\dfrac{2\sin\theta(a\sin\alpha\phi+b\cos\alpha\phi)p_\phi\,p_\theta}{\sin^\alpha\theta\cos^2\theta}\,.\nn
\en
Then  we can introduce auxiliary polynomial
\[
 A(\lambda)=\dfrac{\sin\theta p_\theta}{\alpha\cos\theta}\,\lambda
+\dfrac{a\sin\alpha\phi+b\cos\alpha\phi}{\alpha}\,p_\phi - \dfrac{\sin\theta(a\cos\alpha\phi-b\sin\alpha\phi)}{\alpha\cos\theta}\,p_\theta\,,
\]
such that
\[
\{B(\lambda), A(\mu)\}=\dfrac{1}{\lambda-\mu}\,\Bigl((\mu^2-a^2-b^2)B(\lambda)-(\lambda^2-a^2-b^2)B(\mu)\Bigr)\,,\qquad \{A(\lambda),A(\mu)\}=0\,.
\]
It entails that
\bq\label{p-kow}
p_{j}=-\dfrac{1}{u_j^2-a^2-b^2}\, A(\lambda=q_j)\,,\qquad j=1,2,
\eq
are canonically conjugated momenta on $T^*\mathbb S^2$ with the standard  Poisson brackets
\[ \{q_i,p_j\}=\delta_{ij}\,,\quad \{q_1,q_2\}=\{p_1,p_2\}=0\,.
\]
Below we prove that at  $\alpha=1,2$  this variables are variables of separation for the Kowalevski top and Chaplygin system, respectively.

\subsection{Kowalevski top}
Let us consider Kowalevski top defined by the following integrals of motion
\ben\label{H-kow}
H_1&=&J_1^2+J_2^2+2J_3^2+2bx_1\\
\nn\\
H_2&=&(J_1^2+J_2^2)^2-4b\Bigl(x_1(J_1^2-J_2^2)+2x_2J_1J_2\Bigr)-4b^2x_3^2\,.\nn
\en
In original Kowalevski work \cite{kow89} the first step in the separation of variables method
 is the complexification: she introduces
\[
\mathrm z_1=J_1 + i J_2, \qquad \mathrm z_2=J_1-i J_2
\]
as independent complex variables. Next she makes
her famous change of variables
\[
\label{s1,2} s_{1,2}=\frac{R(\mathrm z_{1},\mathrm z_{2} ) \pm\sqrt{R(\mathrm z_{1},\mathrm z_{1}
)R(\mathrm z_{2},\mathrm z_{2} )}} {2(\mathrm z_{1} -\mathrm z_{2} )^2}.
\]
The fourth degree polynomials $R(\mathrm z_i,\mathrm z_k)$ may be found in \cite{kow89,bm05}.
It brings the system (\ref{eq-mkov}) to the form
\[
(-1)^k\,(s_1-s_2)\dot{s}_k=\sqrt{P(s_k)\,}\,,\qquad
k=1,2,
\]
where
\bq\label{Kow-pol}
P(s)=4\left((s-H_1)^2-\frac{H_2+4b^2C_1}{4}\right)\left[s\left( (s-H_1)^2+b^2C_1-\dfrac{H_2+4a^2C_1}{4}\right)+b^2C_2
\right]\,.
\eq
Consequently, initial equations of motion can be written as hyperelliptic quadratures
\ben
\dfrac{\dot{s}_1}{\sqrt{P(s_1)}}+\dfrac{\dot{s}_2}{\sqrt{P(s_2)}}&=&0\,,\nn\\
\nn\\
\dfrac{s_1\dot{s}_1}{\sqrt{P(s_1)}}+\dfrac{s_2\dot{s}_2}{\sqrt{P(s_2)}}&=&\ii\,,\nn
\en
where we can substitute the conjugated momenta $p_{s_k}$ instead of $\sqrt{P(s_k)}$ in order to get standard Abel-Jacobi form. So, the problem can be integrated in term of genus two hyperelliptic functions of time. Finally,  we have to substitute these functions of time  $s_{k}(t)$ and $p_{s_k}(t)$ into the initial variables $x,J$, the corresponding expressions  may be
found in \cite{kow89,kot93}.

Discussion of the another variables of separation for some particular subcases in  the Kowalevski dynamic may be found in \cite{bm05}.  As usual different variables of separation are related with the distinct integrable  deformations of the  initial integrals of motion.

\subsubsection{New real variables of separation at $C_2=0$}

According to \cite{ts11s}, at $\alpha=1$  coordinates $q_{1,2}$  (\ref{kow-var})  are variables of separation associated with the Hamilton function
\[
H=J_1^2+J_2^2+2J_3^2++2ax_2+2bx_1\,,
\]
which may be reduced to the initial Hamiltonian $H_1$ using rotations (\ref{rot}) around the third axis \cite{kst03},
so we can put $a=0$ in  (\ref{kow-var}) without loss of generality.

Coordinates  $q_{1,2}$  (\ref{kow-var})  at $\alpha=1$ and $a=0$ are defined by
\[
B(\lambda)=(\lambda-q_1)(\lambda-q_2)=\lambda^2+\dfrac{\sqrt{x_1^2+x_2^2}\,(J_1^2+J_2^2)}{x_3^2}
-\dfrac{b\Bigl(x_1(J_1^2-J_2^2)+2x_2J_1J_2\Bigr)}{x_3^2}-b^2\,.
\]
The conjugated   momenta $p_{1,2}$ are equal to
\[
p_k=-\dfrac{A(\lambda=q_k)}{q_k^2-b^2}\,,\qquad
A(\lambda)=\dfrac{x_1J_2-x_2J_1}{x_3}\,\lambda+\dfrac{b\sqrt{x_1^2+x_2^2}J_2}{x_3}\,.
\]
This variables $q_{1,2}$ are differed on a constant terms $\pm b$ from variables introduced in  \cite{ts10k}. Inverse transformation reads as
\ben
x_1&=&\dfrac{b^2-q_1q_2}{b(q_1-q_2)^2}\,\Bigl((b^2-q_1^2)p_1^2+(b^2-q_2^2)p_2^2\Bigr)
-\dfrac{2(b^2-q_1^2)(b^2-q_2^2)}{b(q_1-q_2)^2}\,p_1p_2\,,
\nn\\
\nn\\
x_2&=&-\dfrac{\sqrt{(q_1^2-b^2)(b^2-q_2^2)}}{b(q_1-q_2)^2}\,\Bigl(b^2(p_1-p_2)^2-(p_1q_1-p_2q_2)^2\Bigr)\,,
\nn\\
\nn\\
x_3&=&\sqrt{1-\dfrac{(b^2-q_1^2)^2p_1^4+(b^2-q_2^2)^2p_2^4}{(q_1-q_2)^2}
+\dfrac{2(b^2-q_1^2)(b^2-q_2^2)p_1^2p_2^2}{(q_1-q_2)^2}}\,,
\nn\\
\label{itrans}\\
J_1&=&\dfrac{\sqrt{(q_1^2-b^2)(b^2-q_2^2)}(p_1q_1-p_2q_2)}{(b^2-q_1^2)p_1^2-(b^2-q_2^2)p_2^2}\,\dfrac{x_3}{b}\,,
\nn\\
\nn\\
J_2&=-&\dfrac{q_2(b^2-q_1^2)p_1-q_1(b^2-q_2^2)p_2}{(b^2-q_1^2)p_1^2-(b^2-q_2^2)p_2^2}\,\,\dfrac{x_3}{b}\,,
\nn\\
\nn\\
J_3&=&-\sqrt{(q_1^2-b^2)(b^2-q_2^2)}\,\,\dfrac{p_1-p_2}{q_1-q_2}\,.\nn
\en
Coordinates of separation  take values only in the following intervals
\[ q_1>b>q_2\,,\]
similar to the standard elliptic coordinates on the sphere \cite{bm05}.

In this variables integrals of motion $H_{1,2}$ (\ref{H-kow})  look like
\ben
H_1&=&\dfrac{(b^2-q_1^2)^2p_1^4-(b^2-q_2^2)^2p_2^4-(q_1^2-q_2^2)}{(b^2-q_1^2)p_1^2-(b^2-q_2^2)p_2^2}
\nn\\
\nn\\
H_2&=&\dfrac{\Bigl((b^2-q_1^2)p_1^2-(b^2-q_2^2)p_2^2+q_1+q_2\Bigr)\Bigl((b^2-q_1^2)p_1^2-(b^2-q_2^2)p_2^2-q_1-q_2\Bigr)}{(b^2-q_1^2)p_1^2-(b^2-q_2^2)p_2^2}\,
\nn\\
\nn\\
&\times&\Bigl((b^2-q_1^2)p_1^2-(b^2-q_2^2)p_2^2+q_1-q_2\Bigr)
\Bigl((b^2-q_1^2)p_1^2-(b^2-q_2^2)p_2^2-q_1+q_2\Bigr)\,.\nn
\en
It is easy to see that integrals of motion  and variables of separation are related via the following separated relations
\bq\label{seprel0}
\Phi=\Bigl(2(q^2-b^2)p^2+H_1+\sqrt{H_2}\Bigr)\Bigl(2(q^2-b^2)p^2+H_1-\sqrt{H_2}\Bigr)-4q^2=0,
\eq
at $ q=q_{1,2}$ and $p=p_{1,2}$. Equation  $\Phi(q,p)=0$ defines genus three  hyperelliptic  curve with the following base of the holomorphic differentials
\ben
\Omega_1&=& \dfrac{dq}{p(b^2-q^2)\Bigl(H_1-2(b^2-q^2)p^2\Bigr)}\,,\qquad
\Omega_2= \dfrac{qdq}{p(b^2-q^2)\Bigl(H_1-2(b^2-q^2)p^2\Bigr)}\nn\\
\nn\\
 \Omega_3&=&\dfrac{pdq}{H_1-2(b^2-q^2)p^2},.\nn
\en
In fact  equation (\ref{seprel0}) is invariant with respect to involution $(q,p)\to(-q,p)$. Factorization
with respect to this involution give rise to elliptic curve.

In variables of separation  equations of motion  (\ref{eq-mkov})  have the following  form
\[
\dfrac{\dot{q}_1}{p_1(b^2-q_1^2)\Bigl(H_1-2(b^2-q_1^2)p_1^2\Bigr)}
+\dfrac{\dot{q}_2}{p_2(b^2-q_2^2)\Bigl(H_1-2(b^2-q_2^2)p_2^2\Bigr)}=0\,,
\]
\[
\dfrac{\dot{q}_1}{H_1-2(b^2-q_1^2)p_1^2}+\dfrac{\dot{q}_2}{H_1-2(b^2-q_2^2)p_2^2}=-2\,.
\]
The above quadratures
 in the integral form
\bq\label{ajm-1}
\int^{q_1}_{q_0} \Omega_1+\int^{q_2}_{q_0} \Omega_1=\beta_1\,,\qquad
\int^{q_1}_{q_0} \Omega_3+\int^{q_2}_{q_0} \Omega_3=-2t+\beta_2\,,
\eq
represent the Abel-Jacobi map associated to the genus three hyperelliptic curve defined by $\Phi(q,p)=0$. In particular
it means that instead of $p$ in $\Omega_{1,3}$ (\ref{ajm-1}) we have to substitute  function on $q$ obtained from  the separated relation (\ref{seprel0}).

 In order to give explicit  theta-functions solution  one can apply some remarkable relations between roots of certain functions on symmetric products of such curves and quotients of theta-functions with half-integer characteristics, which are historically referred to as root function and are  generalized   so-called \textit{Wurzelfunktionen} that were used by Jacobi for the case of ordinary hyperelliptic Jacobians \cite{be97,fed99}.
 For the case of odd order hyperelliptic curves such functions were obtained by Weierstrass \cite{w94}.  Inverting the map (\ref{ajm-1}) and substituting symmetric functions of $q_1,q_2,p_1,p_2$ into (\ref{itrans}), one finally finds $x,J$ as functions of time.

\subsubsection{Deformations of the Kowalewski top }
According to \cite{ts10k,ts11s},  using separated relations
\bq\label{seprel1}
\Phi_1=\Bigl(2(q^2-a^2)p^2+\widehat{H}_1+\sqrt{\widehat{H}_2}\Bigr)\Bigl(2(q^2-a^2)p^2+\widehat{H}_1-\sqrt{\widehat{H}_2}\Bigr)-4cu^2+4du+e(q^2-a^2)p=0\,,
\eq
one gets Hamilton function of the generalized Kowalevski top
\bq\label{kow-d1}
\widehat{H}_1=\left(1-\dfrac{c-1}{x_3^2}\right)(J_1^2+J_2^2)+2J_3^2+2bx_1+\dfrac{d}{\sqrt{x_1^2+x_2^2}}+\dfrac{e(x_2 J_1-x_1 J_2)}{4\sqrt{(x_1^2+x_2^2} x_3}\,.
\eq
Second integral of motion is equal to
\ben
\widehat{H}_2&=&\dfrac{(x_3^2+c-1)^2}{x_3^4}\,(J_1^2+J_2^2)^2
-\left(\dfrac{4(x_1J_1^2-x_1J_2^2+2x_2J_1J_2)(x_3^2+c-1)}{x_3^2}+\dfrac{J_2\sqrt{x_1^2+x_2^2}e}{x_3}\right)a\nn\\
\nn\\
&-&4(x_3^2+c-1)a^2+
\left(\dfrac{(2(J_1^2+J_2^2)(c-x_1^2-x_2^2)}{\sqrt{x_1^2+x_2^2}\,x_3^2}+\dfrac{(x_2J_1-x_1J_2)e}{2(x_1^2+x_2^2)x_3}
\right)d
+\dfrac{d^2}{x_1^2+x_2^2}\nn\\
\nn\\
&+&\dfrac{(J_1^2+J_2^2) (x_2 J_1-x_1 J_2) (c-x_1^2-x_2^2) e}{2\sqrt{x_1^2+x_2^2} x3^3}
+\dfrac{(x_2J_1-x_1J_2)^2 e^2}{16 (x_1^2+x_2^2)x_3^2}
\en
According to \cite{kst03,ts11s}  canonical transformation (\ref{g-shift})
reduces Hamilton function (\ref{kow-d1}) to the natural form
\bq\label{kow-h1}
\widehat{H}_1=\left(1-\dfrac{c-1}{x_3^2}\right)(J_1^2+J_2^2)+ 2J_3^2+2ax_1+\dfrac{d}{\sqrt{x_1^2+x_2^2}}
-\dfrac{e^2}{64(x_3^2+c-1)}\,,
\eq
at
\[
f(x_3) =\dfrac{ex_3}{8(x_3^2+c-1)}\,.
\]
At $c=1$ this system coincides with one of the deformations discussed  in \cite{yeh}. Below  we will show only the final form (\ref{kow-h1}) of the deformed  Hamiltonians and will omit the intermediate form (\ref{kow-d1})  for the brevity.

It is easy to calculate the corresponding  equations of motion
\ben
\dfrac{\dot{q}_1}{(b^2-q_1^2)\Bigl(8\widehat{H}_1p_1+e-16p_1^3(b^2-q_1^2)\Bigr)}
+\dfrac{\dot{q}_2}{p_2(b^2-q_2^2)\Bigl(8\widehat{H}_1p_2+e-16p_2^3(b^2-q_2^2)\Bigr)}&=&0\,,
\nn\\
\nn\\
\dfrac{\dot{q}_1}{8\widehat{H}_1p_1+e-16p_1^3(b^2-q_1^2)}+\dfrac{\dot{q}_2}{8\widehat{H}_1p_2+e-16p_2^3(b^2-q_2^2)}&=&-\dfrac{1}{4}\,,
\nn
\en
and prove that  the Abel-Jacobi map on genus three hyperelliptic curve  has the same form (\ref{ajm-1})
\[
\int^{q_1}_{q_0} \Omega_1+\int^{q_2}_{q_0} \Omega_1=\beta_1\,,\qquad
\int^{q_1}_{q_0} \Omega_3+\int^{q_2}_{q_0} \Omega_3=-2t+\beta_2\,,
\]
where  $p$  have to be solution of the separated relation (\ref{seprel1}) and
\[
\Omega_1=\dfrac{dq}{(b^2-q^2)(8\widehat{H}_1p+e-16p^3(b^2-q^2))}\,,\qquad
\Omega_3=\dfrac{dq}{8\widehat{H}_1p+e-16p^3(b^2-q^2)}\,.
\]

\subsection{Chaplygin system}
Let us consider Chaplygin system  defined by the following Hamilton function
\bq\label{chap-hg}
H_1=J_1^2+J_2^2+2J_3^2-2a(x_1^2-x_2^2)-2bx_1x_2-\dfrac{c}{x_3^2}\,.
\eq
At  $c=0$ this system and the corresponding variables of separation have  been investigated by Chaplygin  \cite{ch03}. 
Singular term has been added by Goryachev in \cite{gor16}.

Using rotations (\ref{rot}) around the third axis \cite{kst03} we can put $b=0$ without loss of generality. In this case second integral of motion is equal to
\[
H_2=\left(J_1^2+J_2^2-\dfrac{c}{x_3^2}\right)^2-4ax_3^2(J_1^2-J_2^2)+4a^2x_3^4.
\]
According to \cite{ts10c,ts11s}, coordinates  $q_{1,2}$ (\ref{kow-var})  are variables of separation for this integrable system at
 $\alpha=2$ and $b=0$. In this case   $q_{1,2}$ are roots of the following polynomial  (\ref{kow-var})
\[
B(\lambda)=(\lambda-q_1)(\lambda-q_2)=\lambda^2-\dfrac{J_1^2+J_2^2}{x_3^2}\,\lambda-\dfrac{2aJ_2^2}{x_3^2}+\dfrac{a(J_1^2+J_2^2)}{x_3^2}-a^2\,,
\]
whereas momenta $p_{1,2}$ are values of the other auxiliary polynomial
\[
A(\lambda)=-\dfrac{x_2J_1-x_1J_2}{2x_3}\,\lambda
-\dfrac{ax_1x_2J_3}{x_1^2+x_2^2}-\dfrac{a(x_1^2-x_2^2)(x_2J_1-x_1J_2)}{2(x_1^2+x_2^2)x_3}\,.
\]
at $\lambda=q_{1,2}$ (\ref{p-kow}). Inverse transformation reads as
\ben
x_1&=&\dfrac{ \sqrt{2(q_1-a)(a-q_2)}(p_1(q_1+a)-p_2(q_2+a))}{\sqrt{a}(q_1-q_2)}\,,\nn\\
\nn\\
x_2&=&\dfrac{\sqrt{2(q_1+a)(q_2+a)}(p_1(q_1-a)-p_2(q_2-a))}{\sqrt{a}(q_1-q_2)}\,,\nn\\
\nn\\
x_3&=&\sqrt{1-4\dfrac{(q_1^2-a^2)p_1^2-(q_2^2-a^2)p_2^2 }{q_1-q_2}}\,,\label{inv-ch}\\
\nn\\
J_1&=&\sqrt{\dfrac{(a+q_1)(a+q_2)}{2a}}\,x_3\,,\qquad
J_2=-\sqrt{\dfrac{(q_1-a)(a-q_2)}{2a}}\,x_3\nn\\
\nn\\
J_3&=&-2\sqrt{(q_1^2-a^2)(a^2-q_2^2)}\,\,\dfrac{p_1-p_2}{q_1-q_2}\nn
\en
As usual coordinates of separation  take values only in the following intervals
\[ q_1>a>q_2\,.\]
This variables $q_{1,2}$ are related with variables of separation from \cite{ts10c} by the rule
$q_k\to q_k+a$.

In  variables of separation integrals of motion read as
\ben
H_1&=&4(a^2-q_1^2)p_1^2+4(a^2-q_2^2)p_2^2+q_1+q_2-\dfrac{c}{4(a^2-q_1^2)p_1^2-4(a^2-q_2^2)p_2^2+q_1-q_2}\,,
\nn\\
\nn\\
H_2&=&\Bigl(4(a^2-q_1^2)p_1^2-4(a^2-q_2^2)p_2^2+q_1-q_2\Bigr)^2+\dfrac{c^2(q_1-q_2)^2}{\Bigl(4(a^2-q_1^2)p_1^2-4(a^2-q_2^2)p_2^2+q_1-q_2\Bigr)^2}\nn\\
\nn\\
&-&2c(q_1+q_2)
\,.\nn
\en
It is easy to see that integrals of motion  and variables of separation are related via the following separated relations
\bq\label{seprel-ch}
\Phi=\Bigl(8(q^2-a^2)p^2-2q+H_1-\sqrt{H_2}\Bigr)\Bigl(8(q^2-a^2)p^2-2q+H_1+\sqrt{H_2}\Bigr)-4cq=0,
\eq
at $ q=q_{1,2}$ and $p=p_{1,2}$. Equation  $\Phi(q,p)=0$ defines genus two algebraic curve with the following holomorphic
differentials
\bq\label{h-diff-ch}
\Omega_1= \dfrac{dq}{p(a^2-q^2)\Bigl(H_1-8(a^2-q^2)p^2-2q\Bigr)} \,,\qquad
\Omega_2=\dfrac{\Bigl(4(a^2-q^2)p^2+q\Bigr)dq}{p(a^2-q^2)\Bigl(H_1-8(a^2-q^2)p^2-2q\Bigr)}\,.
\eq
The corresponding  quadratures  look like
\ben
\dfrac{\dot{q}_1}{p_1(a^2-q_1^2)\Bigl(H_1-8(a^2-q_1^2)p_1^2-2q_1\Bigr)} +
\dfrac{\dot{q}_2}{p_2(a^2-q_2^2)\Bigl(H_1-8(a^2-q_2^2)p_2^2-2q_2\Bigr)} &=&0
\nn\\
\nn\\
\dfrac{\Bigl(4(a^2-q_1^2)p_1^2+q_1\Bigr)\dot{q}_1}{p_1(a^2-q_1^2)\Bigl(H_1-8(a^2-q_1^2)p_1^2-2q_1\Bigr)}+
\dfrac{\Bigl(4(a^2-q_2^2)p_2^2+q_2\Bigr)\dot{q}_2}{p_2(a^2-q_2^2)\Bigl(H_1-8(a^2-q_2^2)p_2^2-2q_2\Bigr)}&=&8\,.
\nn
\en
The Abel-Jacobi map on genus two hyperelliptic curve  has the standard  form
\[
\int^{q_1}_{q_0} \Omega_1+\int^{q_2}_{q_0} \Omega_1=\beta_1\,,\qquad
\int^{q_1}_{q_0} \Omega_2+\int^{q_2}_{q_0} \Omega_2=8t+\beta_2\,,
\]
where  $p$ into $\Omega_{1,2}$  have to be solution of the separated relation (\ref{seprel-ch}).

\subsubsection{Deformations of the Chaplygin system}
According to \cite{ts10c,ts11s},  if we substitute this variables of separation onto the following separated relations
\bq\label{seprel-ch1}
\Phi_1=\Bigl(8(q^2-a^2)p^2-2dq+\widehat{H}_1-\sqrt{\widehat{H}_2}\Bigr)
\Bigl(8(q^2-a^2)p^2-2dq+\widehat{H}_1+\sqrt{\widehat{H}_2}\Bigr)-4cq+e(q^2-a^2)p=0\,,
\eq
one gets the Hamilton function of the generalized Chaplygin system
\bq\label{chap-d1}
\widehat{H}_1=\left(1-\dfrac{1-d}{x_3^2}\right)(J_1^2+J_2^2)+2J_3^2-2a(x_1^2-x_2^2)-2bx_1x_2-\dfrac{c}{d-1+x_3^2}+\dfrac{(x_2 J_1-x_1 J_2) e}{8(d-x_1^2-x_2^2) x_3}\,.
\eq
As for the Kowalevski top, using canonical transformation (\ref{g-shift}) at
\[
f(x_3)=\dfrac{ex_3\sqrt{1-x_3^2}}{16(d-1+x_3^2)^2}\]
we can reduce Hamilton function (\ref{chap-d1}) to the natural Hamiltonian
\[
\widehat{H}_1=\left(1-\dfrac{1-d}{x_3^2}\right)(J_1^2+J_2^2)+2J_3^2-2a(x_1^2-x_2^2)-2bx_1x_2-\dfrac{c}{d-1+x_3^2}+
\dfrac{e(x_3^2-1)}{256(d-1+x_3^2)^3}\,.
\]
At $d=1$ additional term is equal to $e(x_3^{-4}-x_3^{-6})$ and this system coincides with one of the deformations considered in \cite{yeh}.

In this case we have genus three hyperelliptic curve with holomorphic differentials
\ben
\Omega_1&=&\dfrac{dq}{(a^2-q^2)\Bigl(e+32p\bigl(\widehat{H}_1-8(a^2-q^2)p^2-2dq\bigr) \Bigr)}\,,\nn\\
\nn\\
\Omega_2&=&\dfrac{qdq}{(a^2-q^2)\Bigl(e+32p\bigl(\widehat{H}_1-8(a^2-q^2)p^2-2dq\bigr) \Bigr)}\,,\nn\\
\nn\\
\Omega_3&=&\dfrac{p^2dq}{e+32p\bigl(\widehat{H}_1-8(a^2-q^2)p^2-2dq\bigr)}\,,\nn
\en
and  the corresponding quadratures  involve  all this differentials
\[
\int^{q_1}_{q_0} \Omega_1+\int^{q_2}_{q_0} \Omega_1=\beta_1\,,\qquad
\int^{q_1}_{q_0} \left(4\Omega_2+d\Omega_3\right)
+\int^{q_2}_{q_0} \left(4\Omega_2+d\Omega_3\right)=-\dfrac{t}4+\beta_2\,,
\]
in contrast with other integrable systems on genus three algebraic curves considered in this note.

\section{Integrable systems associated with trigonal curves}
\setcounter{equation}{0}
 According to \cite{ts09v,ts11s,ts11v},  we introduce  other coordinates  $q_{1,2}$ on $T^8\mathbb S^2$ defined as roots of the following polynomial
\bq\label{q-gor}
B(\lambda)=(\lambda-q_1)(\lambda-q_2)=
\lambda^2-\mathrm i\sqrt{F}\lambda+\Lambda\,,\quad \mathrm i=\sqrt{-1}\,,
\eq
with   coefficients
\bq\label{alp}
 F=\Bigl(g(\theta)p_\theta-\mathrm i h(\theta)p_\phi\Bigr)^2\,,\qquad
 \Lambda=\alpha\,\exp\left(\mathrm i\phi-\int\dfrac{h(\theta)}{g(\theta)}\,d\theta\right)\,,
 \eq
 depending on arbitrary functions $g(\theta)$ and $h(\theta)$. As usual conjugated  momenta $p_{1,2}$ are
 equal to
\bq\label{p-gor}
p_{k}=A(\lambda=q_k)\,,\qquad
 A(\lambda)=\ii\int\dfrac{d\theta}{{g(\theta)}}-\dfrac{\mathrm ip_\phi}{\lambda}\,.
\eq
It is easy to prove, that these polynomials satisfy to the following relations
\bq\label{ab-eq}
\{ B(\lambda), A(\mu)\}=\dfrac{\lambda}{\mu-\lambda}
\left(\dfrac{ B(\lambda)}{\lambda}-\dfrac{ B(\mu)}{\mu}\right)\,,
\qquad \{ A(\lambda), A(\mu)\}=\{ B(\lambda), B(\mu)\}=0\,,\eq
which give rise to canonical Poisson brackets
\[ \{q_i,p_j\}=\delta_{ij}\,,\quad \{q_1,q_2\}=\{p_1,p_2\}=0\,.
\]
 Substituting  variables
\bq\label{subs-a0}
x=a\,q^{-1}_{k}\,,\qquad z= a_0 p_{k}\,,\qquad k=1,2,\qquad a,a_0\in\mathbb R\,,
\eq
into the generic equation of the   (3,4) algebraic curve
\bq\label{g-Phi}
\Phi(z,x)=z^3+(a_1x+a_2)z^2+(H_1x^2+b_1x+b_2)z+x^4+H_2x^3+c_1x^2+c_2x+c_3=0\,,
\eq
 and solving  the resulting equations with respect to $H_{1,2}$,
one gets the following  Hamilton function
\bq\label{g-Ham}
H_1=T+V+\left(
\dfrac{c_2+\ii a_0b_1w_2-a_0^2a_1w_2^2}{a_0aw_2}\,h
+\dfrac{2a_0a_1w_2-\ii b_1}{aw_2}\right)\ii p_\phi-\dfrac{gw_2(c_2+\ii a_0b_1w_2-a_0^2a_1w_2^2)}{aa_0}
p_\theta\,,
\eq
where geodesic Hamiltonian $T$ and potential $V$ are equal to
\ben
T&=&\left(
\dfrac{a_0^2(h^2w_2^2-3hw_2+3)}{a^2}
-\dfrac{\ii a_0a_2(hw_2-1)^2}{a^2w_2}
-\dfrac{b_2h(hw_2-1)}{a^2w_2}+\dfrac{\ii c_3h^2}{a_0a^2w_2}
\right)p_\phi^2\nn\\
\nn\\
&+&\dfrac{\ii g}{a^2w_2}\left(
\Bigl( 2a_0^2w_2^3-2\ii a_0a_2w_2^2-2b_2w2+\dfrac{2\ii c_3}{a_0} \Bigr)h-3a_0^2w_2^2+2\ii a_0a_2w_2+b_2
\right)\,p_\phi p_\theta\nn\\
\nn\\
&+&\dfrac{g^2(a_0b_2w_2+\ii a_0^2a_2w_2^2-a_0^3w_2^3-\ii c_3)}{a^2a_0w_2}\,p_\theta^2\nn\\
\nn\\
V&=&-\dfrac{\ii a^2\mathrm e^{-\ii\phi}}{\alpha a_0 w_1w_2}
+\dfrac{(a_0b_2w_2+\ii a_0^2a_2w_2^2-a_0^3w_2^3-\ii c_3)\alpha w_1\mathrm e^{\ii\phi}}{a_0a^2w_2}
+\dfrac{\ii c_1}{a_0w_2}\,.\nn
\en
Here
\[
w_1=\exp\left(-\int\dfrac{h(\theta )}{g(\theta )}\,d\theta \right) \,,\qquad w_2=\int\dfrac{d\theta }{g(\theta )}\,.
\]
Second integral of motion $H_2$ is a cubic polynomial in momenta $p_{\phi}$ and $p_\theta$.

The resulting  Hamiltonian $H_1$ (\ref{g-Ham}) has the natural form,  if and only if
\[
2a_0a_1w_2-\ii b_1=0\,,\quad
c_2+\ii a_0b_1w_2-a_0^2a_1w_2^2=0
\,.
\]
So,  because $w_2\neq 0$, we have to put
\[
a_1=b_1=c_2=0\,.
\]
 If we want to obtain diagonal metric, then we have to solve integral equation
\bq\label{hg-eq}
2h(a_0^3w_2^3-\ii a_0^2a_2w_2^2-a_0b_2w_2+\ii c_ 3)-3a_0^3w_2^2+2\ii a_0^2a_2w_2+a_0b_2=0\,,
\eq
with respect to functions $h(\theta), w_2(\theta)$ and parameters  $a_0,a_2,b_2,c_3$. If we want to get real potential \[V=f_1(\theta)\cos(\phi)+f_2(\theta)\]
in (\ref{g-Ham}), we have to add one more  equation to (\ref{hg-eq})
 \bq\label{hg-eq2}
\ii\alpha^2(a_0^3w_2^3-\ii a_0^2a_2w_2^2-a_0b_2w_2+\ii c_ 3)w_1^2+a^4=0
 \eq
  depending in addition on function  $w_1$ and  parameters  $a$ (\ref{subs-a0}) and $\alpha$ (\ref{alp}).

Some particular solutions of these equation  have been  studied in  \cite{ts05,ts09v,ts11s}  including  integrable systems due to Goryachev, Chaplygin, Dullin, Matveev etc. For all these systems,  we collect  $a_0$ and the  zero-valued coefficients in (\ref{hg-eq})  in the following table
 \begin{center}
\begin{tabular}{|l|l|l|}
     \hline
  &  & \\
  Goryachev-Chaplygin top & $a_0=2\ii a$ & $b_2=c_3=0$\\
&   & \\
  Goryachev system &$a_0=2\ii a /3$ &$a_2=b_2=0$  \\
 &    &\\
 Case 3 from \cite{ts05} & $a_0=\ii a/3$ &$a_2=b_2=0$  \\
&    &\\
  Dullin-Matveev system &$a_0=\ii a$ &$c_3=0$  \\
&    &\\
 Case 5  from \cite{ts05} &$a_0=\ii a/2$ &$a_2=c_3=0$\\
  \hline
\end{tabular}
\end{center}
Integrable systems with the same coefficients in the separated relations (\ref{g-Phi}) and with
different $a_0$ and $a_0'$ in (\ref{subs-a0}) are related by non-canonical transformation of the momenta
\bq\label{non-tr}
z=a_0p_k\to z=a'_0{p_k}\,, \qquad k=1,2.
\eq

\subsection{Goryachev-Chaplygin top}
Let us consider  Gorychev-Chaplygin top with the following integrals of motion
\bq
H_1=J_1^2+J_2^2+4J_3^2+ax_1+\dfrac{b}{x_3^2}\,,\qquad
H_2= 2J_3\left(J_1^2+J_2^2+\dfrac{b}{x_3^2}\right)+ax_3J_1
\eq
In this case variables of separation   (\ref{q-gor},\ref{p-gor}) are determined by
\[
q_1+q_2=-\dfrac{2J_3}{x_3^2}-\dfrac{J_1+\ii J_2}{x_3(x_1+\ii x_2)}\,,\qquad
q_1q_2=\dfrac{a}{2x_3^2(x_1+\ii x_2)}\,,\qquad
p_{1,2} = \dfrac{\ii x_3^2}{2}+\dfrac{\ii J_3}{q_{1,2}}\,.
\]
They are related with initial variables by the rule
\ben
x_1+\ii x_2&=&-\dfrac{\ii a(q_1-q_2)}{4q_1q_1(p_1q_1-q_2p_2)}\,,\qquad\qquad \qquad x_3=\sqrt{-\dfrac{2\ii(p_1q_1-q_2p_2)}{q_1-q_2}}
\,,\nn\\
\nn\\
J_1+\ii J_2&=&\dfrac{a(q_1^2p_1-q_2^2p_2)}{2q_1q_2(p_1q_1-q_2p_2)\sqrt{-\dfrac{2\ii(p_1q_1-q_2p_2)}{q_1-q_2}}}\,,\qquad J_3=\dfrac{\ii q_1q_2(p_1-p_2)}{q_1-q_2}\,,
\nn\\
\nn\\
x_1-\ii x_2&=&\dfrac{4 q_1q_2}{a(q_1-q_2)^2}\,,
\Bigl(
(\ii-2p_1)q_1^2p_1+(4p_1p_2-\ii p_1-\ii p_2)q_1q_2+(\ii-2p_2)q_2^2p_2
\Bigr)\,,\nn\\
\nn\\
J_1-\ii J_2&=&
-\dfrac{8\ii q_1q_2}{a(q_1-q_2)^2\sqrt{-\dfrac{2\ii(p_1q_1-q_2p_2)}{q_1-q_2}}}
\Bigl
(q_1p_1-q_2p_2)(\ii-2p_1)p_1q_1^2+
\Bigr.\nn\\
\nn\\
&+&\Bigl.(q_1p_1+q_2p_2)(\ii-2p_2)p_2q_2^2\Bigr)\,.\nn
\en
Separated relation is given by equation with real coefficients
\bq\label{seprel-gch1}
\Phi(q,\mu)= (\mu^2-b)q^2+(\mu^3-H_1\mu+H_2)q+\dfrac{a^2}{4}=0\,,\qquad
q=q_{1,2},\quad\mu=2\ii\, q_{1,2}p_{1,2}\,.
\eq
Equations of motion in variables of separation look like
\ben
 \dfrac{\dot{q_1}}{q_1(3\mu_1^2+2q_1\mu_1-H_1)}+
  \dfrac{\dot{q_2}}{q_2(3\mu_2^2+2q_2\mu_2-H_1)}&=&0\,,\nn\\
  \nn\\
 \dfrac{\dot{\mu_1q_1}}{q_1(3\mu_1^2+2q_1\mu_1-H_1)}+
  \dfrac{\dot{\mu_2q_2}}{q_2(3\mu_2^2+2q_2\mu_2-H_1)}&=&2\ii\,.
\nn
\en
By making the birational change
\bq\label{br-change}
q = \dfrac{a^2}{4x},\qquad  \mu =\dfrac{z}{x}
\eq
the curve (\ref{seprel-gch1}) can be transformed to the canonical trigonal form (\ref{g-Phi}) at
\[
a_1=b_1=b_2=c_2=c_3=0\,,
\]
whereas other parameters are functions on $a,b$.

\subsubsection{Deformation of the Goryachev-Chaplygin top}
Substituting $q=q_{1,2}$ and $\mu=2\ii\, q_{1,2}p_{1,2}$ into the
 non-hyperelliptic algebraic curve of genus three defined by the following equation
\bq\label{gch-gyr}
\Phi_1(q,\mu)=cq^3+(\mu^2+d\mu-b)q^2+(\mu^3+e\mu^2-\widehat{H}_1\mu +\widehat{H}_2)q+\dfrac{a^2}4=0\,,
\eq
and solving a pair of the resulting equations with respect to $\widehat{H}_{1,2}$
one gets  deformation of the initial Hamilton function
\ben
\widehat{H}_1&=&J_1^2+J_2^2+4J_3^2+ax_1+\dfrac{b}{x_3^2}-\left( e-\dfrac{c-d+e}{x_1^2+x_2^2}+\dfrac{c}{x_3^2}-\dfrac{2c}{x_3^4} \right)J_3
+\dfrac{(c-dx_3^2+ex_3^4)^2}{4x_3^6(x_1^2+x_2^2)}\,,\nn
\en
using  the generalized shift  (\ref{g-shift}) at
\[
f=-\dfrac{\ii (ex_3^4-dx_3^2+c)}{2\sqrt{1-x_3^2}\,x_3^3}\,.
\]
In this case quadratures are defined by the following differential equations
\ben
\sum_{k=1}^2\dfrac{\dot{q}_k}{q_k(3\mu_k^2+2\mu_kq_k+2e\mu_k+dq_k-\widehat{H}_1)}&=&0\,,
\qquad \mu_k={2\ii}\, q_{k}p_{k}
\nn\\
\nn\\
\sum_{k=1}^2\dfrac{\mu_k\dot{q}_k}{q_k(3\mu_k^2+2\mu_kq_k+2e\mu_k+dq_k-\widehat{H}_1)}&=&{2\ii}\,.\nn
\en
If  $c=0$ and $d=e$,  one gets the usual  Goryachev-Chaplygin gyrostat with the Hamiltonian
\[
\widehat{H}_1=J_1^2+J_2^2+4J_3^2- eJ_3+ax_1+\dfrac{b}{x_3^2}\,.
\]
In this case equation (\ref{gch-gyr}) defines genus two  hyperelliptic curve instead of trigonal one.

 \subsection{Goryachev system}
Let us consider  Gorychev system on the sphere  defined by the following integrals of motion
\ben
H_1&=&J_1^2+J_2^2+\dfrac{4}{3}J_3^2+\dfrac{ax_1}{x_3^{2/3}}+\dfrac{b}{x_3^{2/3}}\,,
\nn\\
\label{gor-int}\\
H_2&=&-\dfrac{2J_3}{3}\left(J_1^2+J_2^2+\dfrac{8}{9}J_3^2+\dfrac{b}{x_3^{2/3}}\right)+\dfrac{a(3x_3J_1-2x_1J_3)}{3x_3^{2/3}}\,.
\nn
\en
The corresponding  variables of separation $q_{1,2}$ and $p_{1,2}$  (\ref{q-gor},\ref{p-gor}) are obtained from
\[
q_1+q_2=\dfrac{x_3^{4/3}J_3}{1-x_3^2}+\dfrac{\ii (J_1x_2-x_1J_2)x_3^{1/3}}{1-x_3^2}\,,\quad
q_1q_2=\dfrac{a}{2(x_1+\ii x_2)}\,,\quad
p_{1,2}= \dfrac{3\ii x_3^{2/3}}{2}+\dfrac{\ii J_3}{q_{1,2}}\,.\quad
\]
Inverse transformation looks like
\ben
x_1+\ii x_2&=&\dfrac{a}{2q_1q_2}\,,\qquad J_3=\dfrac{\ii q_1q_2(p_1-p_2)}{q_1-q_2}\,,
\nn\\
\nn\\
x_1-\ii x_2&=& 2\dfrac{q_1q_2(1-x_3^2)}{a}\,,\qquad J_1+\ii J_2= -\dfrac{a(q_1+q_2)}{2q_1q_2x_3^{1/3}}\,,\nn\\
\nn\\
J_1-\ii J_2&=& -\dfrac{4\ii q_1^2q_2^2(p_1-p_2)}{a(q_1-q_2)}\,x_3
+\dfrac{2q_1q_2(1-x_3^2)(q_1+q_2)}{ax_3^{1/3}}\,,
\en
where
\[
x_3=\left(-\dfrac{2\ii(p_1q_1-p_2q_2)}{3(q_1-q_2)}\right)^{3/2}\,.
\]
Separated relation is given by equation with real coefficients
\bq\label{seprel-gor}
\Phi(q,\mu)= q^4-bq^2+(\mu^3-H_1\mu+H_2)q+\dfrac{a^2}{4}=0\,,\qquad\mbox{at}\quad
q=q_{1,2},\quad\mu=\dfrac{2\ii}3\, q_{1,2}p_{1,2}\,.
\eq
In this case  quadratures read as
\ben
\int_{q_0}^{q_1}\dfrac{dq}{q(3\mu^2-H_1)}+\int_{q_0}^{q_2}\dfrac{dq}{q(3\mu^2-H_1)}&=&\beta_1\,,\nn\\
\label{aj-gor1}\\
\int_{q_0}^{q_1}\dfrac{\mu dq}{q(3\mu^2-H_1)}+\int_{q_0}^{q_2}\dfrac{\mu dq}{q(3\mu^2-H_1)}&
=&\dfrac{2\ii}{3}\,t+\beta_2\,.\nn
\en
 As usual, here  $\mu$ is a  function on $q$ obtained from  the separated relation (\ref{seprel-gor}).

\subsubsection{Deformation of the Goryachev  system}
Using  trigonal  curve of genus three defined by the following equation
\bq\label{gor-gyr}
\Phi_1=q^4+cq^3-bq^2+(\mu^3+d\mu^2-\widehat{H}_1\mu+\widehat{H}_2)q+\dfrac{a^2}{4}=0\,,
\eq
instead of (\ref{seprel-gor})  one gets  deformation of the initial Hamilton function (\ref{gor-int})
\ben
\widehat{H}_1&=&H_1-\left(\dfrac{d}{3}+\dfrac{d+cx_3^{2/3}}{x_1^2+x_2^2}  \right)\,J_3
+\dfrac{(c+dx_3^{4/3})^2}{4(x_1^2+x_2^2)x_3^{2/3}}
\en
after the generalized shift  (\ref{g-shift}) at
\[
f=-\dfrac{\ii (c+dx_3^{4/3})}{2\sqrt{1-x_3^2} x_3^{1/3}}\,.
\]
The corresponding equations of motion  look like
\ben
\dfrac{\dot{q}_1}{q_1(3\mu_1^2+2d\mu_1-\widehat{H}_1)}+\dfrac{\dot{q}_2}{q_2(3\mu_2^2+2d\mu_2-\widehat{H}_1)}&=&0\,,
\qquad \mu_k=\dfrac{2\ii}3\, q_{k}p_{k}
\nn\\
\nn\\
\dfrac{\mu_1\dot{q}_1}{q_1(3\mu_1^2+2d\mu_1-\widehat{H}_1)}
+\dfrac{\mu_1\dot{q}_2}{q_2(3\mu_2^2+2d\mu_2-\widehat{H}_1)}&=&\dfrac{2\ii}{3}\,.\nn
\en
Birational transformation (\ref{br-change}) maps
the curve (\ref{gor-gyr})  to the canonical trigonal form (\ref{g-Phi}) at $a_2=b_1=b_2=0$\,.

\subsection{Case 3 from \cite{ts05}}
Let us consider one more  integrable system from  \cite{ts05} defined by the following integrals of motion
\ben
H_1&=&J_1^2+J_2^2+\left(\dfrac{1}{12}+\dfrac{(2x_3+1)}{2(x_3+1)}\right)J_3^2
+\dfrac{ax_1}{(x_3+1)^{5/6}}+\dfrac{b}{(x_3+1)^{1/3}}\,,
\nn\\
\label{ts1-int}\\
H_2&=&\dfrac{1}{27}\,J_3^3-\dfrac{1}{3}\,J_3H_1-a(x_3+1)^{1/6}\,J_1+\dfrac{ax_1J_3}{2(x_3+1)^{5/6}}\,.
\nn
\en
The corresponding  variables of separation $q_{1,2}$ and $p_{1,2}$  (\ref{q-gor},\ref{p-gor}) are obtained from
\[
q_1+q_2=-\dfrac{(1+x_3)^{2/3}J_3}{2(x_3-1)}-\dfrac{\ii(x_2J_1-x_1J_2)}{(1+x_3)^{1/3}(x_3-1)}\,,\quad
q_1q_2=\dfrac{a\sqrt{1+x_3}}{2(x_1+\ii x_2)}\,,\quad
p_{1,2}=3\ii(1+x_3)^{1/3}+ \dfrac{\ii J_3}{q_{1,2}}\,.\quad
\]
Inverse transformation looks like
\ben
x_1+\ii x_2&=&\dfrac{a\sqrt{1+x_3}}{2q_1q_2}\,,\qquad J_3=\dfrac{\ii q_1q_2(p_1-p_2)}{q_1-q_2}\,,
\nn\\
\nn\\
x_1-\ii x_2&=& -\dfrac{2q_1q_2(x_3^2-1)}{a\sqrt{1+x_3}}\,,\qquad J_1+\ii J_2=\dfrac{\ii a(p_1-p_2)}{4(q_1-q_2)\sqrt{1+x_3}}-\dfrac{a(q_1+q_2)}{2q_1q_2(1+x_3)^{1/6}}\,,\nn\\
\nn\\
J_1-\ii J_2&=&-\dfrac{\ii q_1^2q_2^2(3x_3+1)(p_1-p_2)}{a(q_1-q_2)\sqrt{1+x_3}}-\dfrac{2q_1q_2(q_1+q_2)(x_3-1)}{a(1+x_3)^{1/6}}\,,
\en
where
\[
x_3=\dfrac{\ii(p_1q_1-p_2q_2)^3}{27(q_1-q_2)^3}-1\,.
\]
Separated relations are defined by equation with the real coefficients
\bq\label{seprel-ts1}
\Phi(q,\mu)=2q^4-bq^2+(\mu^3\,q-H_1\mu+H_2)q+\dfrac{a^2}{4}=0\,,\qquad
q=q_{1,2},\quad \mu=\dfrac{\ii\, q_{1,2}p_{1,2}}{3}
\eq
The corresponding quadratures are given by
\ben
\int_{q_0}^{q_1}\dfrac{\dot{q}}{q(3\mu^2-H_1)}+\int_{q_0}^{q_2}\dfrac{\dot{q}}{q(3\mu^2-H_1)}&=&\beta_1\,,\nn\\
\label{aj-31}\\
\int_{q_0}^{q_1}\dfrac{\mu\dot{q}}{q(3\mu^2-H_1)}+\int_{q_0}^{q_2}\dfrac{\mu\dot{q}}{q(3\mu^2-H_1)}&=&\dfrac{\ii}{3}\, t+\beta_2\,.\nn
\en
As for the  Goryachev system,  birational change (\ref{br-change}) transforms
the equation  (\ref{seprel-ts1})  to the canonical trigonal form (\ref{g-Phi}) at
\[a_1=a_2=b_1=b_2=c_2=0\,,\]
 It allows us to prove that  integrals of motion  for this system (\ref{ts1-int}) are related with  integrals of motion (\ref{gor-int}) for the  Goryachev system by the non-canonical transformation (\ref{non-tr}).

 It may seem that quadratures (\ref{aj-gor1}) and (\ref{aj-31}) are trivially related by  change of time
 \[t\to2t\,,\]
 but we have to keep firmly in mind that $\mu$ in (\ref{aj-gor1})  is a function  on $q$ obtained from  (\ref{seprel-gor}),  whereas $\mu$ in (\ref{aj-31}) is another function on $q$ obtained from  (\ref{seprel-ts1}).

 \subsubsection{Deformation of the system (\ref{ts1-int}) }
Similar to  the Goryachev system, we can add two term to the initial trigonal  curve of genus three  (\ref{seprel-ts1})
\bq\label{ts1-gyr}
\Phi_1=2q^4+cq^3-bq^2+(\mu^3+d\mu^2-\widehat{H}_1\mu+\widehat{H}_2)q+\dfrac{a^2}{4}=0\,.
\eq
Deformation of the initial Hamilton function (\ref{ts1-int}) looks like
\ben
\widehat{H}_1&=&H_1-\left(\dfrac{d}{6}-\dfrac{d}{x_3-1}-\dfrac{c(1+x_3)^{1/3}}{2(x_3-1)}  \right)\,J_3
+\dfrac{\left(d\sqrt{1+x_3}+c(1+x_3)^{-1/6}\right)^2}{4(1-x_3)}
\en
after canonical transformation  (\ref{g-shift}) at
\[
f= -\dfrac{\ii\Bigl(d(1+x_3)+c(1+x_3)^{1/3}\Bigr)}{2\sqrt{1-x_3^2}}\,.
\]
In this case equations of motion are equal to
 \ben
\dfrac{\dot{q}_1}{q_1(3\mu_1^2+2d\mu_1-\widehat{H}_1)}+\dfrac{\dot{q}_2}{q_2(3\mu_2^2+2d\mu_2-\widehat{H}_1)}&=&0\,,\qquad
\mu_k=\dfrac{\ii}3\, q_{k}p_{k}\,, \nn\\
\nn\\
\dfrac{\mu_1\dot{q}_1}{q_1(3\mu_1^2+2d\mu_1-\widehat{H}_1)}
+\dfrac{\mu_2\dot{q}_2}{q_2(3\mu_2^2+2d\mu_2-\widehat{H}_1)}
&=&\dfrac{\ii}{3}\,.
\en

\subsection{Dullin-Matveev system}
Let us consider the Dullin-Matveev system \cite{dull04} defined by the following integrals of motion
\ben
H_1&=&J_1^2+J_2^2+\left(1+\dfrac{x_3}{x_3+c}-\dfrac{x_3^2-|x|^2}{4(x_3+c)^2}\right)
J_3^2+\dfrac{ax_1}{(x_3+c)^{1/2}}+\dfrac{b}{x_3+c}\,,\nn\\
\label{int-dm}\\
H_2&=& -\left(J_1^2+J_2^2-\dfrac{J_3^2}{4}+\dfrac{(4x_3^2+6x_3c+c^2+|x|^2)J_3^2}{4(x_3+c)^2}+\dfrac{b}{x_3+c}\right)J_3\nn\\
\nn\\
&+&a\sqrt{x_3+c}J_1-\dfrac{ax_1J_3}{2\sqrt{x_3+c}}\,.\nn
\en
According to \cite{ts11s} variables of separation $q_{1,2}$ and $p_{1,2}$ are defined by (\ref{q-gor},\ref{p-gor})
\[
q_1+q_2=-\dfrac{J_3}{2(c+x_3)}-\dfrac{J_1+\ii J_2}{x_1+\ii x_2}\,,\qquad
q_1q_2=\dfrac{a}{2(x_1+\ii x_2)\sqrt{c+x_3}}\,,\qquad
p_{1,2}= \ii(c+x_3)+\dfrac{\ii J_3}{q_{1,2}}
\]
or by inverse transformation
\ben
x_1+\ii x_2&=&\dfrac{a}{2\sqrt{-\dfrac{\ii (p_1q_1-p_2q_2)}{q_1-q_2}}q_1q_2}\,,\quad
x_3=-\dfrac{\ii (p_1q_1-q_2p_2)}{q_1-q_2}-c\,,\quad J_3=\dfrac{\ii q_1q_2(p_1-p_2)}{q_1-q_2}\,,
\nn\\
\nn\\
x_1-\ii x_2&=&-\dfrac{
2\sqrt{-\dfrac{\ii (p_1q_1-p_2q_2)}{q_1-q_2}}\,q_1q_2
}{a(q_1-q_2)^2}\,\Bigl((c+1+\ii p_1)q_1-(c+1+\ii p_2)q_2\Bigr)\nn\\
\nn\\
&\times& \Bigl((c-1+\ii p_1)q_1-(c-1+\ii p_2)q_2\Bigr)\,,
\nn\\
\nn\\
J_1+\ii J_2&=&-\dfrac{a(q_1(2q_1+q_2)p_1-q_2(2q_2+q_1)p_2)}{4\sqrt{-\dfrac{\ii (p_1q_1-p_2q_2)}{q_1-q_2}}\,,q_1q_2(p_1q_1-p_2q_2)}\,,
\label{inv-dm}\\
\nn\\
J_1-\ii J_2&=&\dfrac{\ii q_1q_2}{a\sqrt{-\dfrac{\ii (p_1q_1-p_2q_2)}{q_1-q_2}} (q_1-q_2)}\,
\Bigr(
(c+1+\ii p_1)(c-1+\ii p_1)(2p_1q_1-3q_2p_1-p_2q_2)q_1^3\Bigl.\nn\\
\nn\\
&+&\bigl(2\ii (p_1+p_2)c-4p_1p_2\bigr)(p_1-p_2)q_1^2q_2^2
-(c+1+\ii p_2)(c-1+\ii p_2)(2p_2q_2-3q_1p_2-p_1q_1)q_2^3\,.\nn
\en
The corresponding separated relations are defined by equation with the real coefficients
\bq\label{seprel-dm}
\Phi(q,\mu)=\mu(c^2-1)q^3+(2c\mu^2-b)q^2+(\mu^3-H_1\mu+H_2)\,q+\dfrac{a^2}{4}=0\,,\qquad
q=q_{1,2},\quad \mu=\ii\, q_{1,2}p_{1,2}\,,
\eq
and quadratures in differential form look like
\ben
 \dfrac{\dot{q_1}}{q_1\Bigl((c^2-1)q_1^2+4cq_1\mu_1+3\mu_1^2-H_1\Bigr)}+
  \dfrac{\dot{q_2}}{q_2\Bigl((c^2-1)q_2^2+4cq_2\mu_2+3\mu_2^2-H_1\Bigr)}&=&0\,,\nn\\
  \nn\\
\dfrac{\mu_{1}\dot{q_1}}{q_1\Bigl((c^2-1)q_1^2+4cq_1\mu_1+3\mu_1^2-H_1\Bigr)}+
\dfrac{\mu_{2}\dot{q_2}}{q_2\Bigl((c^2-1)q_2^2+4cq_2\mu_2+3\mu_2^2-H_1\Bigr)}&=&\ii\,.\nn
\en

\subsubsection{Deformation of the Dullin-Matveev system}
Substituting $q=q_{1,2}$ and $\mu=\ii\, q_{1,2}p_{1,2}$ into the
 non-hyperelliptic algebraic curve of genus three defined by the following equation
\bq\label{dm-gyr}
\Phi_1=\mu(c^2-1)q^3+(2c\mu^2+d\mu-b)q^2+(\mu^3+e\mu^2-\widehat{H}_1\mu+\widehat{H}_2)q+\dfrac{a^2}{4}=0\,,
\eq
and solving a pair of the resulting equations with respect to $H_{1,2}$
one gets  deformation of the initial Hamilton function (\ref{int-dm})
\ben
\widehat{H}_1&=&H_1-\dfrac{1}{2}\left( e-\dfrac{d}{c+x_3}+\dfrac{(ce-d)x_3+e}{x_1^2+x_2^2}  \right)J_3+
-\dfrac{(ce-d+x_3e)^2}{4(x_1^2+x_2^2)}
\en
after the generalized shift  (\ref{g-shift}) at
\[
f=-\dfrac{\ii (ce-d+x_3e)}{2\sqrt{1-x_3^2}}\,.
\]
Using the same birational change (\ref{br-change})
the curve (\ref{gor-gyr}) can be transformed  to the canonical trigonal form (\ref{g-Phi}) at $c_2=c_3=0$.

In this case equations of motion read as
\ben
\sum_{k=1}^2\dfrac{\dot{q}_k}{q_k(3\mu_k^2+4cq_k\mu_k+2e\mu_k+q_k^2(c^2-1)+dq_k-\widehat{H}_1)}&=&0\,,
\qquad \mu_k=\ii q_kp_k\,,
\nn\\
\label{dmd-q}\\
\sum_{k=1}^2\dfrac{\mu_k\dot{q}_k}{q_k(3\mu_k^2+4cq_k\mu_k+2e\mu_k+q_k^2(c^2-1)+dq_k-\widehat{H}_1)}
&=&{\ii}\,.\nn
\en

\subsection{Case 5 from \cite{ts05}}
Let us consider last  integrable system from  \cite{ts05} with integrals of motion
\ben
H_1&=&J_1^2+J_2^2+\left(\dfrac{3}{16}+\dfrac{8x_3+5}{8(x_3+1)}\right)\,J_3^2+
\dfrac{ax_1}{(x_3+1)^{3/4}}+\dfrac{b}{\sqrt{x_3+1}}\,,
\nn\\
\label{ts2-int}\\
H_2&=&\dfrac{1}{8}\,J_3^3-\dfrac{1}{2} H_1J_3+a(x_3+1)^{1/4}J_1--\dfrac{ax_1J_3}{4(x_3+1)^{3/4}}\,.
\nn
\en
The corresponding  variables of separation $q_{1,2}$ and $p_{1,2}$  (\ref{q-gor},\ref{p-gor}) are obtained from
\[
q_1+q_2=\dfrac{(3x_3+1)J_3}{4\sqrt{x_3+1}(1-x_3)}+\dfrac{\ii(x_2J_1-x_1J_2)}{\sqrt{x_3+1}(1-x_3)}\,,\quad
q_1q_2=\dfrac{a(x_3+1)^{1/4}}{2(x_1+\ii x_2)}\,,\quad
p_{1,2}=2\ii\sqrt{x_3+1}+\dfrac{\ii J_3}{q_{1,2}}\,.\quad
\]
Inverse transformation looks like
\ben
x_1+\ii x_2&=&\dfrac{a(x_3+1)^{1/4}}{2q_1q_2}\,,\qquad J_3=\dfrac{\ii q_1q_2(p_1-p_2)}{q_1-q_2}\,,
\nn\\
\nn\\
x_1-\ii x_2&=& -\dfrac{2q_1q_2(x_3^2-1)}{a(x_3+1)^{1/4}}\,,\qquad J_1+\ii J_2=
\dfrac{\ii a(p_1-p_2)}{8(q_1-q_2)(x_3+1)^{3/4}}-\dfrac{a(q_1+q_2)}{2q_1q_2(x_3+1)^{1/4}}
\,,\nn\\
\nn\\
J_1-\ii J_2&=&-\dfrac{\ii q_1^2q_2^2(7x_3+1)(p_1-p_2)}{2a(q_1-q_2((x_3+1)^{1/4}}-\dfrac{2q_1q_2(q_1+q_2)(x_3-1)(x_3+1)^{1/4}}{a}\,,
\en
where
\[
x_3=-\dfrac{(p_1q_1-p_2q_2)^2}{4(q_1-q_2)^2}-1\,.
\]
Separated relations are defined by
\bq\label{seprel-ts2}
\Phi(q,\mu)=-2\mu q^3-bq^2+(\mu^3-H_1\mu+H_2)q+\dfrac{a^2}{4}=0\,,\qquad
q=q_{1,2},\quad \mu=\dfrac{\ii}{2}\, q_{1,2}p_{1,2}
\eq
and we have the following  quadratures in differential form
\ben
\dfrac{\dot{q}_1}{q_1(3\mu_1^2-H_1-2q_1^2)}+\dfrac{\dot{q}_2}{q_2(3\mu_2^2-H_1-2q_2^2)}&=&0\,,\nn\\
\nn\\
\dfrac{\mu_1\dot{q}_1}{q_1(3\mu_1^2-H_1-2q_1^2)}+\dfrac{\mu_2\dot{q}_2}{q_2(3\mu_2^2-H_1-2q_2^2)}&=&\dfrac{\ii}{2}\,.
\en

 \subsubsection{Deformation of the system (\ref{ts2-int}) }
Let us add three terms to  initial trigonal  curve of genus three  (\ref{seprel-ts2})
\bq\label{ts2-gyr}
\Phi_1=(c-2 \mu) q^3-(d\mu+b) q^2+(\mu^3+e\mu^2-\widehat{H}_1\mu +\widehat{H}_2) q+\dfrac{a^2}{4}=0\,.
\eq
The corresponding  deformation of the initial Hamilton function (\ref{ts2-int}) has the form
\ben
\widehat{H}_1&=&H_1-\left(\dfrac{e}{4}+\dfrac{c+2e}{2(1-x_3)}+\dfrac{c}{4(1+x_3)}
+\dfrac{d(x_3^2+4x_3+3)}{4(1-x_3)(1+x_3)^{3/2}}
\right)\,J_3\nn\\ \nn\\
&+&\dfrac{1}{4(1-x_3)}\left(e\sqrt{1+x_3}+d+\dfrac{c}{\sqrt{1+x_3}}\right)^2\,,
\en
after canonical transformation  (\ref{g-shift}) at
\[
f=-\dfrac{\ii c+\ii e (1+x_3)}{2\sqrt{1-x_3^2}}-\dfrac{\ii d}{2\sqrt{1-x_3}}\,.
\]
The corresponding quadratures are defined by
 \ben
\sum_{k=1}^2\dfrac{\dot{q}_k}{q_k(3\mu_k^2+2e\mu_k-2q_k^2-dq_k-\widehat{H}_1)}
&=&0\,,\qquad
\mu_k=\dfrac{\ii}2\, q_{k}p_{k}\,, \nn\\
\label{c5d-q}\\
\sum_{k=1}^2\dfrac{\mu_k\dot{q}_k}{q_k(3\mu_k^2+2e\mu_k-2q_k^2-dq_k-\widehat{H}_1)}
&=&\dfrac{\ii}{2}\,.\nn
\en
Non canonical transformations (\ref{non-tr}) relate this equations (\ref{c5d-q}) with
similar equations  (\ref{dmd-q})  for the deformed Dullin-Matveev system.

\section{Conclusion}
In  \cite{ts10c,ts10k,ts09v}  some new variables of separation for various integrable systems on the sphere with higher order integrals of motion have been obtained by brute force method. In \cite{ts10,ts11s} we introduce  a concept of natural Poisson, which allows us to understand the geometric origin of this method and to find some common attributes of the variables of separation for the Kowalevski top, Chaplygin system, Goryachev-Chaplygin gyrostat, Goryachev and  Dulllin-Matveev systems etc.

In this more  technical paper we continue our investigations in order to explicitly describe  canonical transformations of  initial physical variables  to  variables of separation and vice versa, to calculate the corresponding quadratures and to discuss possible integrable  deformations of these systems associated with genus three hyperelliptic and non-hyperelliptic algebraic curves.

In Section 2 we consider  real variables of separation for which  the separation relations have the real coefficients only. In Section 3 we discuss  complex variables of separation and  the separation relations with the real coefficients as above.
Similar complex  variables satisfying to the real separated equations for the Kowalevski top and Goryachev-Chaplygin gyrostat 
have been found in â \cite{kuz},  for the Kowalevski-Goryachev-Chaplygin gyrostat in  \cite{ts02} and for the Steklov-Lyapunov system in  \cite{ts11sl}.  These and other known complex variables lying on the real algebraic curves  are discussed in the book  \cite{bm05}.

Further inquiry is related with numerical, algebro-geometric and topological analysis  of the obtained quadratures.  For dynamical  systems associated with the (3,4) trigonal curve (\ref{g-Phi}) we also want to discuss an application of the Kowalevski-Painleve criteria to these systems, because in generic case solutions of the corresponding quadratures are non-meromorphic functions of time.

We would like to thank A.V. Borisov and Yu.N. Fedorov  for helpful  discussions.


\begin{thebibliography}{10}

\bibitem{bbe94}
E.D. Belokolos, A.I. Bobenko, V.Z. Enolskii, A.R. Its, V.B. Matveev,
 \newblock{\em Algebro-geometrical approach to nonlinear integrable equations} Springer Series in Nonlinear Dynamics, Berlin: Springer-Verlag 1994, XII+320 p.

\bibitem{bm05}
A.V. Borisov, I.S. Mamaev,
\newblock{\em Rigid Body Dynamics. Hamiltonian Methods, Integrability, Chaos}, Moscow-Izhevsk, RCD, 2005.

\bibitem{be97}
 V. M. Buchstaber, V. Z. Enolskii,  D. V. Leikin, \newblock{\em Kleinian functions, hyperelliptic Jacobians
and applications}, Amer. Math. Soc. Transl., Ser. 2, Vol. 179, Amer. Math. Soc., Providence,
RI, 1997, pp. 1–33.

\bibitem{ch03}
S.A. Chaplygin, \newblock{\em  A new partial solution of the problem of motion of a rigid body
in liquid}, Trudy Otdel. Fiz. Nauk Obsh. Liub. Est., v.11, no. 2, p.7–10, 1903.



\bibitem{dull04}
H.R. Dullin, V.S. Matveev,
\newblock{\em A new integrable system on the sphere},
Mathematical Research Letters, v.11, p.715-722, 2004.


\bibitem{fed99}
 Yu. Fedorov, \newblock{\em Classical integrable systems related to generalized Jacobians}, Acta Appl. Math.,
v.55, n. 3, p. 151–201, 1999.

\bibitem{gor15} 
D.N. Goryachev,
\newblock{\em New cases of a rigid body motion about a fixed point,}
Warshav. Univ. Izv., v.3, p.1-11, 1915.

\bibitem{gor16}
D.N. Goryachev,\newblock{\em  New cases of integrability of Euler's dynamical equations},Warsaw
Univ. Izv., v.3, p. 1–13, 1916.

\bibitem{tsg05}
Yu.A. Grigoryev, A.V.Tsiganov,
\newblock{\em Symbolic software for separation of variables in the Hamilton-Jacobi equation for the L-systems},
Regular and Chaotic Dynamics, v.10(4), p.413-422, 2005.


\bibitem{jac42}
C.G. Jacobi.,
\newblock{\em Vorlesungen \"{u}ber Dynamik}, K\"{o}nigsberg University 1842 -
1843 (edited by Clebsch and published from Reimer, Berlin, 1884)

 \bibitem{kst03} I.V. Komarov, V.V. Sokolov, A.V. Tsiganov,
 \newblock{\em Poisson maps and integrable deformations of Kowalevski top.},
 \newblock{ J. Phys. A.}, v.36, p. 8035-8048, 2003.

\bibitem{kot93}
F. Kotter,
 \newblock{\em Sur le cas trait\'{e} par Mme Kowalevski de rotation d'un corps solide pesant autor d'un point fixe},
 Acta Mathematica. v.17, n.1-2. p. 209-263,  1893.

\bibitem{kow89}
S. Kowalevski,
\newblock{\em Sur le probl\'{e}me de la rotation d'un corps solide
autour d'un point fixe},
\newblock{\em Acta Math.}, \textbf{12}, 177-232, 1889.

\bibitem{kuz}
V. B. Kuznetsov
\newblock{\em Simultaneous separation for the Kowalevski and Goryachev–Chaplygin gyrostats},
 J. Phys. A: Math. Gen. v.35, p.6419 , 2002.



\bibitem{li55}
J. Liouville, \newblock{\em Note sur les ´equations de la dynamique}, J. Math. Pures Appl.
v.20, p. 137-138, 1855.

\bibitem{neu59}
C. Neumann,
\newblock{\em De problemate quodam mechanico, quod ad primam integralium
ultraellipticorum classem revocatur}, J. Reine Angew.Math.,  v.56, p. 46-63, 1859.

\bibitem{st95}
P. St\"{a}ckel,
\newblock{\em \"{U}ber die Integralen der Hamilton-Jacobischen Differential Gleichung
mittelst Separation der Variable}, Habilitationsschrift, Halle, 1891.

 \bibitem{ts02}
A.V. Tsiganov,	\newblock{\em On the Kowalevski-Goryachev-Chaplygin gyrostat},
J. Phys. A, Math. Gen. v.35, No.26, L309-L318, 2002.

\bibitem{ts05}  A.V. Tsiganov,
\newblock{\em On a family of integrable systems on ${\mathcal S}^2$ with a cubic integral of motion},
J. Phys. A, Math. Gen. v.38, p.921-927, 2005.


\bibitem{ts07}  A.V. Tsiganov,
\newblock{\em On the two different bi-Hamiltonian structures for the Toda lattice},
J. Phys. A: Math. Theor. v.40, pp. 6395-6406, 2007


\bibitem{ts10c}
A.V. Tsiganov,
\newblock{\em On the generalized Chaplygin system},
Journal of Mathematical Sciences, v.168, n.8, p.901-911, 2010.

\bibitem{ts10k}
A.V. Tsiganov, \newblock{\em New variables of separation for particular case of the Kowalevski top},
Regular and Chaotic Dynamics, v.15, n.6, p. 657-667, 2010.

\bibitem{ts10} A.~V.~Tsiganov, \newblock{\em On bi-integrable natural
    Hamiltonian systems on the Riemannian manifolds}, arXiv: 1006.3914, accepted to Journal of Nonlinear Mathematical Physics,   2010.

\bibitem{ts11s} A.~V.~Tsiganov, \newblock{\em On natural Poisson bivectors on the sphere}, 	J. Phys. A: Math. Theor., v.44, 105203 (15pp), 2011.
    
\bibitem{ts11sl} A.~V.~Tsiganov, \newblock{\em New variables of separation for the Steklov--Lyapunov system},
Preprint: 	arXiv:1101.4345v1, 2011.

\bibitem{ts09v}
A.V. Vershilov, A.V. Tsiganov,
\newblock{\em
On bi-Hamiltonian geometry of some integrable systems on the sphere with cubic integral of motion}, J. Phys. A: Math. Theor. v.42, 105203 (12pp), 2009.

\bibitem{ts11v}
A.V. Vershilov, A.V. Tsiganov,
\newblock{\em
On one integrable system with a cubic first integral}, Preprint 	arXiv:1103.1444v1,  2011.

\bibitem{w94}  K.  Weierstrass,\newblock{\em Mathematische Werke I}, vol. 1, 1894.

\bibitem{yeh}
H.M. Yehia, A.A. Elmandouh,\newblock{\em New integrable systems with a quartic integral and new generalizations of Kovalevskaya's and Goriatchev's cases},
Regular and Chaotic Dynamics, v.13(1), pp. 56 - 69, 2008.

\end{thebibliography}
\end{document}